\def\ArxivVersion{1}

\ifdefined\ArxivVersion
  \documentclass[amsmath]{aastex702}
\else
  \documentclass[linenumbers,amsmath]{aastex702}
\fi
\usepackage{bm}
\usepackage{tikz}
\usepackage{tabularx}
\usepackage{booktabs}
\usepackage{placeins}
\IfFileExists{xurl.sty}{\usepackage{xurl}}{}
\submitjournal{The Astronomical Journal}

\newif\ifshowfigsource
\showfigsourcefalse

\newcommand{\figsource}[1]{\ifshowfigsource\par\smallskip\noindent{\footnotesize Source program: \texttt{\detokenize{#1}}.}\fi}

\newcommand{\AodTriMedianPointKm}{5.7}

\newcommand{\AodTriFitRmsKm}{5.1}

\newcommand{\BeverlyPathRcsTxGainDb}{7.0}
\newcommand{\BeverlyPathRcsRxGainDb}{3.0}

\newcommand{\OrbitTrajectoryFitVersion}{0.3.1}
\newcommand{\OrbitSampleCount}{2000}

\newcommand{\OrbitTopNeaName}{2024 XR16}
\newcommand{\OrbitTopNeaDH}{0.093}

\newcommand{\OrbitTopCometName}{P/2009 WX51 (Catalina)}
\newcommand{\OrbitTopCometDH}{0.308}

\newcommand{\OrbitTopStreamName}{omicron-Eridanids}
\newcommand{\OrbitTopStreamDH}{0.212}

\newcommand{\OrbitTopStreamRadiantSepDeg}{48}

\newcommand{\CorrFitNHeadRange}{229}
\newcommand{\CorrFitNHeadDoppler}{137}
\newcommand{\CorrFitNTrail}{146}

\newcommand{\CorrFitReducedChiSq}{1.29}

\begin{document}

\shorttitle{Radar and infrasound analysis of the Cape Cod Bay bolide}
\shortauthors{Vierinen et al.}

\title{Radar and Infrasound Observations of the 2026 May 30 Cape Cod Bay Superbolide}
\correspondingauthor{Juha Vierinen}

\author[orcid=0000-0001-7651-708X]{Juha Vierinen}
\affiliation{Department of Physics and Technology, UiT - The Arctic University of Norway, Tromsø, Norway}
\affiliation{Leibniz Institute of Atmospheric Physics at the University of Rostock, 18225 Kuehlungsborn, Germany}
\email[show]{jvi019@uit.no}

\author[orcid=0000-0002-7504-0336]{Ryan Volz}
\affiliation{Haystack Observatory, Massachusetts Institute of Technology, 99 Millstone Road, Westford, MA 01886, USA}
\email{rvolz@mit.edu}

\author[orcid=0000-0003-2761-6692]{Marc Fries}
\affiliation{Unaffiliated}
\email{mfries8@hotmail.com}

\author[orcid=0000-0003-4778-1409]{Elizabeth A. Silber}
\affiliation{Sandia National Laboratories, Albuquerque, NM, USA}
\email{esilbe@sandia.gov}

\author[orcid=0000-0002-2369-1997]{Sarah Albert}
\affiliation{Sandia National Laboratories, Albuquerque, NM, USA}
\email{salber@sandia.gov}

\author[orcid=0000-0002-0031-9324]{Philip J. Erickson}
\affiliation{Haystack Observatory, Massachusetts Institute of Technology, 99 Millstone Road, Westford, MA 01886, USA}
\email{pje@mit.edu}

\author[orcid=0000-0002-0118-8223]{Matthias Clahsen}
\affiliation{Leibniz Institute of Atmospheric Physics at the University of Rostock, 18225 Kuehlungsborn, Germany}
\email{clahsen@iap-kborn.de}

\author[orcid=0000-0002-9097-4380]{Frank D. Lind}
\affiliation{Haystack Observatory, Massachusetts Institute of Technology, 99 Millstone Road, Westford, MA 01886, USA}
\email{flind@mit.edu}

\author{Nico Pfeffer}
\affiliation{Leibniz Institute of Atmospheric Physics at the University of Rostock, 18225 Kuehlungsborn, Germany}
\email{pfeffer@iap-kborn.de}

\author[orcid=0000-0003-2627-2031]{John Swoboda}
\affiliation{Haystack Observatory, Massachusetts Institute of Technology, 99 Millstone Road, Westford, MA 01886, USA}
\email{swoboj@mit.edu}

\author[orcid=0000-0003-4732-358X]{Allan Weatherwax}
\affiliation{Haystack Observatory, Massachusetts Institute of Technology, 99 Millstone Road, Westford, MA 01886, USA}
\email{wxwax@mit.edu}

\author[orcid=0000-0003-1992-4118]{Simon G. Shepherd}
\affiliation{Thayer School of Engineering, Dartmouth College, Hanover, NH, USA}
\email{simon.g.shepherd@dartmouth.edu}

\author[orcid=0000-0002-2364-8892]{Jorge L. Chau}
\affiliation{Leibniz Institute of Atmospheric Physics at the University of Rostock, 18225 Kuehlungsborn, Germany}
\email{chau@iap-kborn.org}

\begin{abstract}
Metre-scale impactors connect near-Earth-object dynamics, atmospheric-entry physics and meteorite delivery, but their pre-impact orbits and material properties are rarely measured together, especially for daylight events. We report observations of the 30 May 2026 Cape Cod Bay bolide. A very high frequency (VHF) meteor radar network detected atmospheric-entry plasma, weather radars detected falling meteorites, and infrasound sensors recorded the associated hypervelocity shock. VHF radar range--Doppler measurements resolve the meteor head echo from about 90 to 71 km altitude, while interferometric measurements localize short-lived non-specular plasma wake echoes down to an unusually low altitude of \(\sim\!30\) km. We identify this as the radar terminus, where breakup is expected to occur. A combined head-echo and trail fit constrains a steep southeastward trajectory over coastal Massachusetts. Weather-radar detections refine the dark-flight solution, place the likely fall region in Cape Cod Bay and constrain probable fragment sizes. Backward propagation of the fitted trajectory gives a heliocentric orbit near the cometary--asteroidal classification boundaries, with conflicting classifications from empirical dynamical criteria. Catalogue comparisons identify no convincing parent body or established meteor-shower association. Regional infrasound independently supports the terminal fragmentation region and yields an energy estimate of 0.5 kt TNT equivalent. The radar and dark-flight evidence are consistent with a dense, mechanically strong meteoroid, possibly iron-rich. The combination of meteor-radar plasma tracking, weather-radar observations of falling meteorites and infrasound measurements can provide accurate orbital elements and trace large meteoroids from atmospheric entry to nearly ground impact.
\end{abstract}


\section{Introduction}

Detailed measurements of metre-scale hypervelocity atmospheric entries are
rare \citep{brown_2016_orbital}. Observations of their trajectories, plasma wakes, fragmentation, and
energy release are needed to test and improve atmospheric-entry models in
this extreme high-Mach-number regime \citep{Silber2018}. For planetary defense, accumulating
pre-impact orbits of these high-energy impactors is also essential to
characterizing their source populations and encounter conditions
\citep{brown_2016_orbital,jenniskens_devillepoix_2025}.

On 2026 May 30, a bright daytime bolide crossed the Massachusetts coast and
ended over Cape Cod Bay.  Public accounts reported a widely heard sonic boom,
indicating a meteoroid large enough to generate a shock front. The NASA/JPL Center for Near-Earth Object Studies (CNEOS) reported a U.S. Government sensor detection, with peak brightness at 18:06:23 UTC near \(42.0^\circ\) N, \(70.5^\circ\) W and 32.0 km altitude, a total radiated
energy of \(4.21\times10^{11}\) J, and an estimated impact energy of 1.1 kt of trinitrotoluene (TNT) equivalent \citep{nasa_cneos_2026_fireballs}. Only three events in the CNEOS catalogue for 2025 had estimated impact energies exceeding the 1.1 kt reported for the Cape Cod Bay bolide.  For scale, the reported impact energy is comparable to the 1.2 kt TNT-equivalent asteroid 2008 TC3 impact over Sudan and about one quarter of the 4 kt
Sutter's Mill event, both reported by \citet{jenniskens_2012_sutters_mill}. This places
the Cape Cod Bay bolide in the metre-scale impactor regime.%

\begin{figure}
  \centering
  \includegraphics[width=\linewidth]{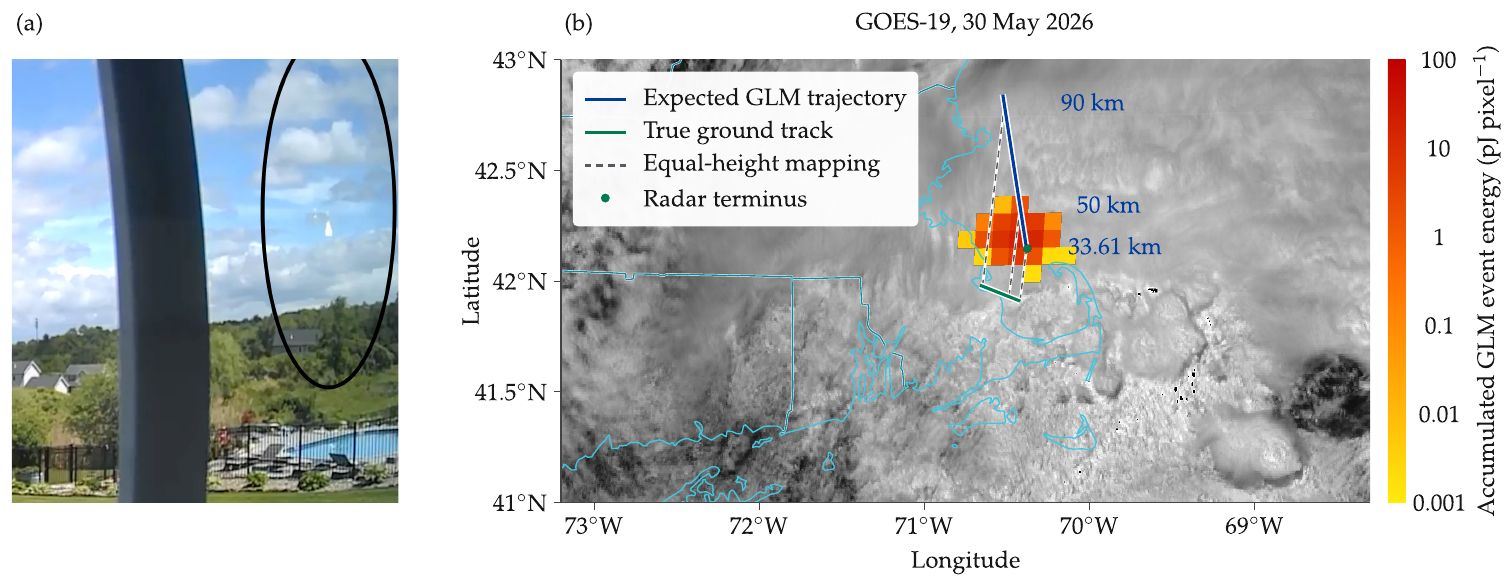}
  \caption{Optical observations of the Cape Cod Bay bolide on 30 May 2026.
  (a) Daytime image from
  R. Schott's WYZE security camera in Mechanicville, New York.
  The black ellipse identifies the fireball.
  Frame counting relative to the camera timestamp transitions gives
  18:06:22.28 UTC, with an adopted timing uncertainty of $\pm50$ ms,
  assuming an accurately synchronized camera clock.
  The fitted trajectory places the meteoroid at $40.0\pm1.2$ km altitude.
  Image credit: \copyright{} R. Schott,
  via the American Meteor Society
  fireball-event data set, event 3867-2026 \citep{ams_2026_event3867}.
  (b) GOES-19 GLM Level 2 event radiant energy at the instrument.  The blue line shows the fitted three-dimensional trajectory projected onto
  the GLM lightning ellipsoid, and the green line shows its true ground track estimated from radar observations.
  Dashed lines connect corresponding positions at 90, 50, and 33.61 km altitude;
  the green dot marks the projected 33.61 km radar terminus.
  Data: NOAA/NESDIS.}
  \figsource{reports/extract_keynote_optical_panels.py; reports/prepare_goes_glm_flash_density.py; reports/rust_goes/src/main.rs}
  \label{fig:doorbell-optical}
\end{figure}

Serendipitously, this event was captured by a wide range of sensors that could be used to track the meteoroid from atmospheric entry down to near sea level and elucidate the nature of the meteoroid: dashcam and security cameras recorded the path through openings in clouds, infrasound sensors recorded the acoustic signature generated by the shock wave, the Massachusetts Institute of Technology (MIT) Haystack Observatory specular meteor radar network detected echoes from the atmospheric entry plasma in the upper atmosphere, and four weather radars detected signatures of falling meteorites in the lower atmosphere. Together, these observations constrain the orbit, provide a detailed description of the atmospheric entry, and predict the meteorite strewn field.
Figure~\ref{fig:doorbell-optical} shows: (a) the bolide in a security-camera frame
from Mechanicville, New York, and (b) GOES-19 Geostationary Lightning Mapper
(GLM) observations. The location of the camera is shown in Figure \ref{fig:station-ballistic-overview}.
The GLM optical transient was  reported by CIMSS
\citep{cimss_2026_goes19_glm}; the use of GLM observations for bolide detection
is discussed by \citet{smith_2026_finding}.

We combine meteor-radar, weather-radar, and infrasound observations to
reconstruct the trajectory and pre-impact orbit, constrain fragmentation and
energy release, and connect atmospheric entry to the predicted meteorite
strewn field. We examine the unusually low-altitude plasma echoes and assess
what the observations imply about the body's mass, mechanical strength, and
possible composition.

\section{Methods}

\begin{figure}
  \centering
  \begingroup
  \normalsize
  \dimen0=\linewidth
  \advance\dimen0 by -164.600000mm
  \ifdim\dimen0<0pt
    \PackageError{rustplotlib}{Article linewidth changed}{Regenerate this figure using the new ArticleLayout.}
  \fi
  \dimen0=\csname f@size\endcsname pt
  \advance\dimen0 by -10.000000pt
  \ifdim\dimen0<0pt \dimen0=-\dimen0\fi
  \ifdim\dimen0>0.01pt
    \PackageError{rustplotlib}{Article body font changed}{Regenerate this figure using the new ArticleLayout.}
  \fi
  \includegraphics[width=128.400000mm]{\detokenize{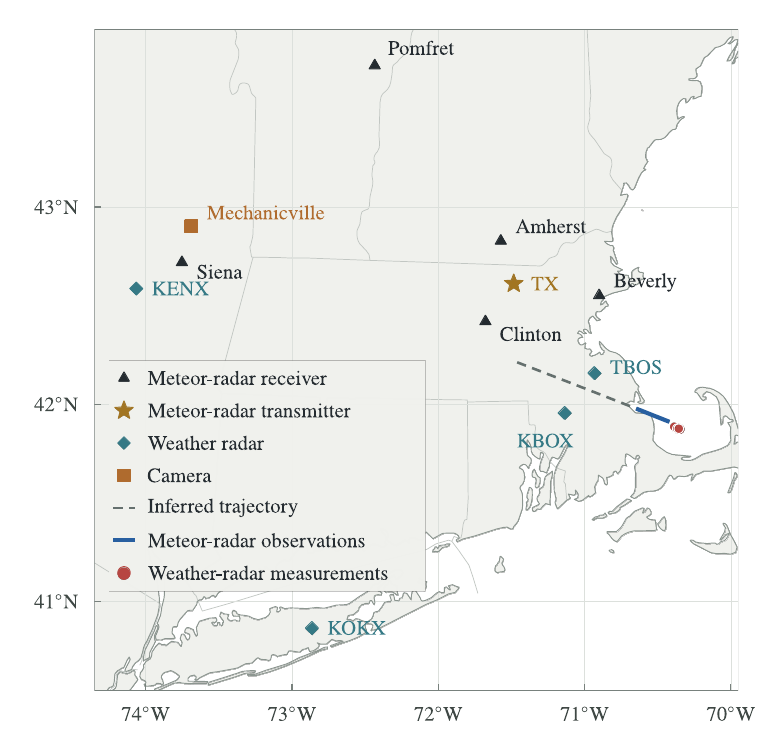}}
  \endgroup
  \caption{Overview of the observing geometry and bolide trajectory.  The
  yellow star marks the MIT Haystack transmitter, black triangles mark the
  five VHF meteor-radar receiving stations, and teal diamonds mark the KBOX,
  KENX, KOKX, and TBOS weather radars. The orange square marks the
  location of the Mechanicville camera.
  The VHF system observes entry-generated
  plasma, whereas the weather radars detect descending solid fragments. Red
  dots mark the observed weather-radar measurement positions.
  The dashed gray curve shows the inferred bolide trajectory
  extrapolated up to 300 km altitude.
  The blue segment indicates the portion of the trajectory containing
  SIMONe/VHF radar measurements starting at 90 km altitude and ending at $\approx$ 30 km altitude.}
  \figsource{reports/plot_article_observing_geometry.py}
  \label{fig:station-ballistic-overview}
\end{figure}

The primary trajectory data are obtained from raw complex voltages \citep{digital_rf_2_6_14,volz_2026_20507006} of the 31.65 MHz MIT Haystack Observatory 
Spread Spectrum Interferometric Meteor Observing Network (SIMONe) coded continuous-wave transmitters and meteor-radar network receivers
\citep{vierinen_2016_mimo,chau_2019_mimo,chau_2021_peru}.  The radar system employs five linearly polarized transmit antennas radiating 500 W each. The five channels utilize different pseudorandom 10 ms long continuous binary phase coded transmit waveforms with 100 $\mu$s bit lengths, allowing estimation of transmit-target-receiver propagation path-length, rate-of-change of path length, as well as interferometric angle-of-departure \citep{chau_2019_mimo}. The radar receivers employ dual polarized antennae located in Beverly, Clinton, Pomfret, Siena University, and Amherst (Figure~\ref{fig:station-ballistic-overview}).
At the VHF frequency used by the radar, the measured radar echoes arise from scattering by the plasma produced by the atmospheric entry of the meteoroid and plasma turbulence in its wake -- not direct detections of solid meteoroids.

In contrast, the microwave weather radars, with a much shorter wavelength, are sensitive to small solid meteoroids in dark flight. This is possible because the meteoroids are comparable in size to the radar wavelength, allowing for resonant scattering.  We used volume-scan data from the NOAA S-band WSR-88D weather radars
KBOX, KENX, and KOKX, and from the FAA Doppler weather
radar TBOS, which operates at C band. Radar-signature centroids and times
constrain fragment descent. Figure~\ref{fig:station-ballistic-overview}
shows the weather-radar sites and the measured falling meteorite positions.

Infrasound measurements provide independent constraints on the bolide's
source direction and energy through the atmospheric pressure waves generated
during entry and fragmentation. We surveyed stations within 2000 km of the
CNEOS peak-brightness location, obtaining usable records from 199 sensors,
including conventional microbarometers, the Elginfield Infrasound Array
(ELFO), and citizen-operated Raspberry Shake/Boom pressure sensors.
Figure~\ref{fig:infrasound_map} shows the infrasound station locations.
Candidate arrivals were tested against acoustic travel-time windows and
local background noise; array beamforming and inter-station envelope delays
constrained the arrival direction, while dominant signal periods provided
an empirical energy estimate. %

\subsection{Meteor radar}
\label{sec:aod}
\label{sec:combined-ballistic-fit}

The SIMONe Haystack meteor radar raw voltage data were processed using range-Doppler matched filtering for the high-Doppler-shift meteor head echoes for the plasma surrounding the moving meteoroid, and compressive sensing radar signal processing \citep{urco2019sparse} for the slowly moving plasma wake. The head-echo analysis estimated the transmitter--target--receiver path lengths and their rates of change. We used a matched filter to detect the five transmit signals and searched for the best-matching angle of departure (AoD) and polarization to maximize the signal-to-noise ratio and estimate an independent interferometric location for the head echoes.

The slowly moving, range-spread plasma structures formed in the meteoroid's wake were analyzed using a compressive sensing approach, which simultaneously deconvolves the continuous coded transmit signals from the five transmitters \citep{urco_2019_maarsy}. This allowed three-dimensional localization of the meteor trail plasma turbulence using interferometric angles of departure and propagation delays \citep{urco_2018_coherent_mimo}.

The head-echo radar cross section (RCS) was estimated using the bistatic radar equation:
\begin{equation}
  P_r =
  \frac{P_t G_t G_r \lambda^2 \sigma_{\mathrm{RCS}}}
       {(4\pi)^3 R_t^2 R_r^2},
  \label{eq:bistatic-radar-equation}
\end{equation}
Here, $P_r$ is the received signal power, $P_t$ is the transmit power of one
transmit antenna, $R_t$ is the transmitter--target distance, and $R_r$ is the
target--receiver distance. The effective transmit and receive gains toward
the target, $G_t$ and $G_r$, are linear power ratios, and $\lambda$ is the
radar wavelength. The effective bistatic RCS, $\sigma_{\mathrm{RCS}}$, is
expressed in square metres and corresponds to the transmitted linear-polarization
state and the polarization-matched combination of the two receive channels.

The single mode noise power is
\begin{equation}
  P_n=k_B T_{\mathrm{sys}}B,
  \label{eq:rcs-noise-power}
\end{equation}
where $k_B$ is Boltzmann's constant, $T_{\mathrm{sys}}$ is the system noise
temperature, and $B$ is the matched-filter noise bandwidth. We used
$B=100~\mathrm{Hz}$ for a single 10 ms code. At the radar frequency of
31.65 MHz ($\lambda=9.47~\mathrm{m}$), sky noise,
primarily from Galactic synchrotron emission, dominates receiver noise.
We therefore used $T_{\mathrm{sys}}=1.54\times10^4~\mathrm{K}$, obtained
from the LWA1 low-frequency sky model
\citep{dowell_2017_lwa1_lfss,price_2016_pygdsm}, giving
$P_n=2.12\times10^{-17}~\mathrm{W}$.

Defining the linear signal-to-noise power ratio as
$\gamma_{\mathrm{SNR}}=P_r/P_n$, the RCS estimator becomes
\begin{equation}
  \sigma_{\mathrm{RCS}} =
  \frac{\gamma_{\mathrm{SNR}}\,k_B T_{\mathrm{sys}}B(4\pi)^3R_t^2R_r^2}
       {P_tG_tG_r\lambda^2}.
  \label{eq:rcs-from-snr}
\end{equation}
We evaluated $R_t$ and $R_r$ at each sample using the meteoroid trajectory.
The conversion assumes $P_t=500~\mathrm{W}$, an effective transmit gain of
$\BeverlyPathRcsTxGainDb~\mathrm{dB}$ for trajectory-directed transmit combining,
and an effective receive gain of $\BeverlyPathRcsRxGainDb~\mathrm{dB}$ for the
polarization-matched receiver. The power and gain values
are first-order estimates rather than absolute calibrations, so the absolute
RCS should be regarded as a rough order-of-magnitude estimate.

The trajectory was fitted in the Geocentric Celestial Reference System (GCRS) by
combining timed head-echo range/Doppler measurements with non-specular trail
positions. In the fit, the head echoes constrain position and velocity at
the time of observation. The trail echoes, on the other hand, provide
point-on-path constraints: the measured trail-echo positions constrain the
meteoroid's path, but not the precise times at which it passed those positions,
because the echoes originate from the turbulent plasma wake left behind by
the meteoroid. The trajectory model included atmospheric drag, Earth's
gravitational acceleration, and Earth's rotation. The atmospheric density
was taken from an NRLMSIS~2.1 altitude profile \citep{emmert2022nrlmsis}.

Because of the object's large mass and the relatively short head-echo path
between altitudes of 90 and 70 km, the measurements could not constrain
the deceleration due to atmospheric drag. NASA's Meteoroid Environment Office
estimated a diameter of 1.6 m and a mass of 5600 kg
\citep{nasa_ares_2026_cape_cod_bay}. For the fixed-drag calculation, we adopted
a 1.6 m diameter sphere with a mass of 5000 kg and a drag coefficient of one,
giving a fixed drag-area-to-mass ratio of
\(4.021\times10^{-4}\ \mathrm{m^2\,kg^{-1}}\). The trajectory fit is not
sensitive to this choice, as no observable deceleration could be detected
from the head echo.

Head-echo range and Doppler measurements were selected independently by
manual inspection of the station range--Doppler products. The fit retained
\CorrFitNHeadRange{} ranges, \CorrFitNHeadDoppler{} Doppler measurements,
and \CorrFitNTrail{} trail positions. Trail selections required SNR
\(\geq8\) dB, altitudes of 20--115 km, and an AoD offset within
\(35^\circ\) of the search prior; an isolated Beverly AoD solution at
18:06:24.88 UTC was excluded.

The six inertial-state parameters were estimated by generalized least squares
using head-echo range and Doppler residuals and three-dimensional trail
residuals to the closest point on the propagated path. Head-echo error scales
were estimated from residual scatter, giving \(\sigma_R=828\) m and
\(\sigma_f=25\) Hz; no SNR dependence was retained. Trail uncertainties used
the interferometric three-dimensional position error, bounded to 1.2--12 km,
as an isotropic per-coordinate standard deviation. Different stations and
observable components were treated as independent covariance blocks. The
three-code incoherent head-echo averages, stepped by one 10 ms code, gave
within-block correlations of \(2/3\) and \(1/3\) at lags of one and two
codes. Trail correlations followed the overlap of eight-code coherent
integrations and were restricted to the same station and raw 3 km range gate;
no retained trail pairs met both conditions. The state covariance was
\((\bm J^{\mathsf T}\bm\Sigma^{-1}\bm J)^{-1}\), where \(\bm J\) is the
residual Jacobian and \(\bm\Sigma\) is the measurement covariance. The
reduced \(\chi^2\) was \CorrFitReducedChiSq{}.

The fitted inertial state and covariance were sampled and transformed using
the DE440 planetary ephemeris \citep{park_2021_planetary}. Samples were propagated backward with the
REBOUND \(N\)-body integrator \citep{rein_liu_2012_rebound} until they were outside
Earth's sphere of influence, then converted to heliocentric osculating
Keplerian elements referred to the mean ecliptic and equinox of J2000.0.

\subsection{Weather radar and dark-flight modeling}

The J{\"o}rmungandr2 (J2) dark-flight model combined the SIMONe path with the
weather-radar signature centroids to define the post-fireball flight regime
\citep{fries_fries_2010,fries_2023_jormungandr}.  The 30--34 km plasma-detection range set the luminous-flight
terminus; each weather-radar centroid was back-propagated to 3 km s\(^{-1}\) to
estimate a common terminus.  J2 then iteratively solved for the fragment mass
that traversed the terminus-to-detection distance in the observed time under
the measured winds. The wind profile was constructed from a distance-weighted
average of NOAA radiosonde measurements from the three stations nearest the
fall site. J2 also extrapolated each fragment to sea level and
calculated an extended strewn field for generic masses from 0.1 g to 10 kg.
Total fall mass was estimated both by assuming 10\% survival of the
pre-atmospheric mass and by integrating the J2 radar-derived mass--number
distribution; the latter is a currently unvalidated method presented for
informational purposes only.

\subsection{Infrasound Observations and Analysis}
\label{sec:infrasound-methods}

Infrasound observations were used as an independent constraint on the source direction and energy of the Cape Cod Bay bolide. The reference source location was the peak-brightness location reported by the National Aeronautics and Space Administration/Jet Propulsion Laboratory Center for Near-Earth Object Studies (NASA/JPL CNEOS; 42.0$^\circ$ N, 70.5$^\circ$ W; 32.0 km altitude), for which a calculated impact energy of 1.1 kilotons of trinitrotoluene (TNT) equivalent was reported (1 kt TNT = $4.184\times10^{12}$ J). The VHF radar terminus at 41.9105$^\circ$ N, 70.4202$^\circ$ W and 33.61 km altitude was retained as an independent spatial reference.

All openly available infrasound stations within 2000 km of the CNEOS location were surveyed. Usable waveform data were obtained from 199 sensors, including conventional single-station microbarometers, the four-element Elginfield Infrasound Array (ELFO) of Natural Resources Canada, and citizen-operated Raspberry Shake/Boom pressure sensors. All waveform data were accessed through the International Federation of Digital Seismograph Networks (FDSN).

Candidate single-station arrivals were restricted to physically admissible acoustic propagation windows following established bolide-infrasound screening practice \citep{Silber2024,Wilson2025}. Arrival windows were defined using nominal tropospheric, stratospheric, and thermospheric celerity ranges \citep{Negraru2010}. To quantify the likelihood of coincident background signals, we developed a matched-window empirical false-alarm test in which the event window was compared with equal-duration off-source windows from the same station record. The empirical false-alarm probability was defined as the fraction of off-source windows whose peak envelope amplitude equaled or exceeded that of the candidate arrival. Candidates were subsequently inspected in the filtered waveform and spectrogram, and ambiguous arrivals were excluded from quantitative analysis.

ELFO was analyzed using sliding-window frequency-wavenumber (FK) beamforming to estimate back-azimuth and trace velocity, following standard infrasound array-processing principles \citep{Rost2002,Costley2013}. Element records were decimated to 25 Hz and bandpass-filtered from 0.14 to 2.0 Hz, with the passband selected from the array signal-to-noise spectrum. Bartlett FK processing was performed in 25 s windows advanced by 5 s over a propagation-slowness grid spanning $\pm3.6$ s/km in 0.05 s/km increments. For each window, time-domain semblance was calculated at the preferred FK slowness after alignment, demeaning, and root-mean-square normalization of the individual element records, providing an amplitude-normalized measure of inter-element waveform coherence \citep{NeidellTaner1971}. Signal onset was defined as the first sustained, source-consistent window exceeding the pre-arrival semblance background at a physically admissible celerity \citep{Silber2024}. Directional uncertainty was estimated by element jackknife applied to an independent cross-correlation delay solution \citep{SzuberlaOlson2004,Green2015}.

A spatially distributed cluster of six Raspberry Shake/Boom sensors, located 65--101 km from the source and separated by up to 36 km, was analyzed jointly as an array of opportunity. Because the inter-station separations substantially exceeded the acoustic wavelength of the observed signal, conventional phase-coherent FK processing was not appropriate. Inter-station delays were instead estimated from waveform-envelope correlations, and the preferred plane-wave solution was obtained by maximizing envelope coherence over back-azimuth and trace velocity. Detection significance was evaluated against randomized time-shift and off-source coherence distributions, and directional uncertainty was estimated by station jackknife. This approach extends previous use of spatially distributed single-sensor stations for bolide localization through multilateration \citep{Wilson2025} by obtaining a direct direction-of-arrival constraint from inter-station coherence.

Dominant signal periods were measured at maximum signal amplitude using the established zero-crossing approach for bolide infrasound, with spectral-peak periods used as an independent consistency check \citep{Ens2012,ReVelle1997,Silber2024}. Source energy was estimated from the network-averaged dominant period using the bolide-specific multi-station period-yield relation of \citet{Silber2025},
\begin{equation}
\log_{10}\left(E\,[\mathrm{kt}]\right)
=
3.71\,\log_{10}\left(P\,[\mathrm{s}]\right)-2.07,
\end{equation}
where $E$ is the source energy in kilotons of TNT equivalent and $P$ is the network-averaged dominant period in seconds. The energy estimate used the five retained single-station detections together with the ELFO beam period. 

\section{Results}

\subsection{VHF radar echoes}

\begin{figure}
  \centering
  \includegraphics[width=\linewidth]{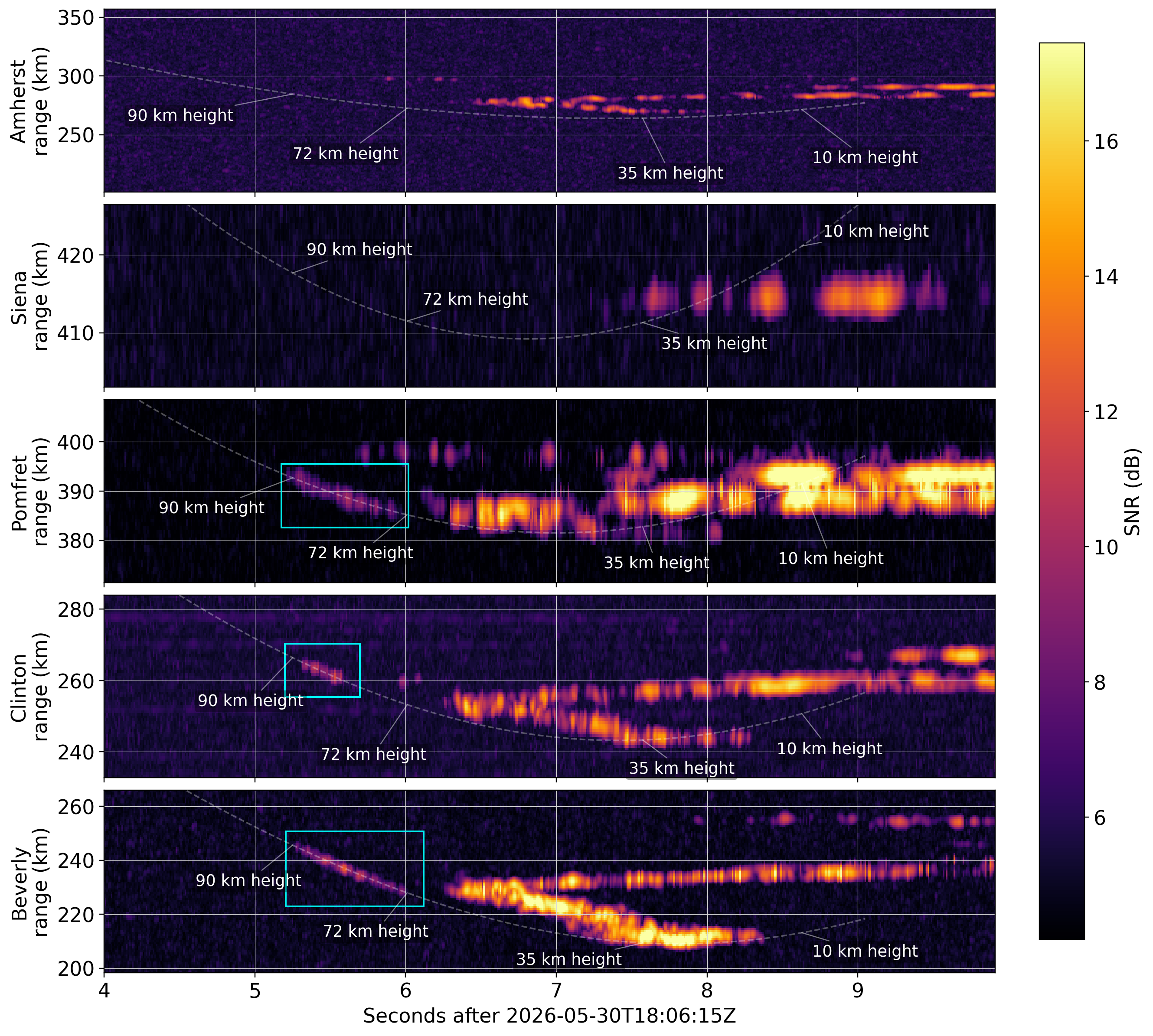}
  \caption{Doppler-whitened signal-to-noise ratio at the SIMONe VHF meteor
  radar receivers in Amherst, Siena, Pomfret, Clinton, and Beverly.
  Range--Doppler matched filtering used eight coherently integrated 10 ms
  codes (80 ms), giving a Doppler resolution of 12.5 Hz, with echo powers averaged
  over the five transmit codes and two receive channels. The plotted SNR is based on the
  strongest-bin power over $-50$ to $+50$ kHz Doppler range.  The faint dashed white curves indicate the
  object's bistatic range predicted by the estimated meteoroid trajectory,
  with annotations indicating the altitude of the meteoroid.
  Cyan outlines mark the regions where meteor head echoes were detected.}
  \figsource{trail_echo_roi.py}
  \label{fig:trail-echo-roi}
\end{figure}

Figure~\ref{fig:trail-echo-roi} shows the range--time--echo power structure on all five VHF
links. Range here is the total transmit-target-receiver propagation distance.  The high-Doppler head echo is visible inside the cyan selection boxes
at Beverly, Clinton, and Pomfret and follows the decreasing bistatic range of
the fitted trajectory from approximately 90 to 71 km altitude.  Clinton
contains the shortest retained head-echo segment; Amherst and Siena did not detect head echoes.
The brighter, more range-extended returns that follow are non-specular plasma
trail echoes.  At altitudes above approximately 72 km their onset is delayed
by several seconds relative to passage of the body, whereas at lower altitudes
the trail response develops progressively closer to the trajectory passage
time.  Trail echoes remain detectable far below the head-echo interval, down to approximately 30 km.

\begin{figure}
  \centering
  \includegraphics[width=\linewidth]{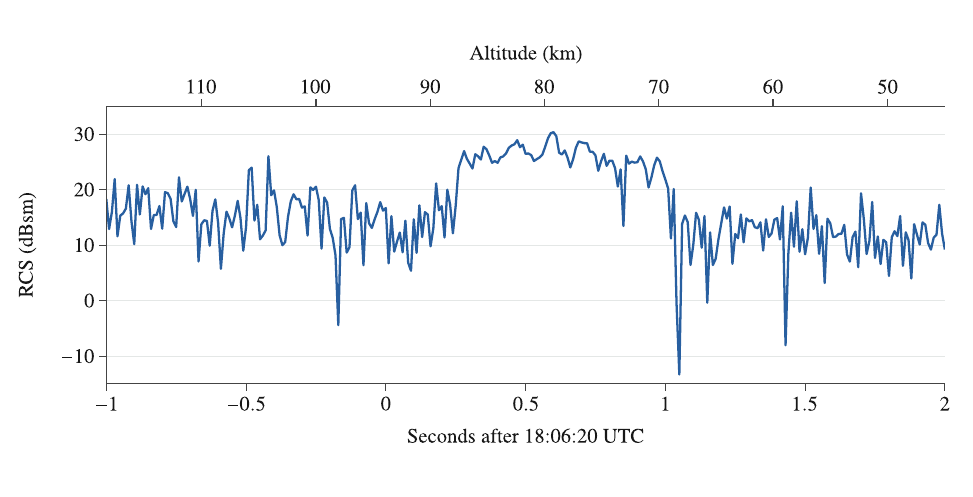}
  \caption{Beverly effective bistatic
  head-echo RCS estimated with range, Doppler, angle-of-departure, and
  polarization matching.}
  \figsource{reports/render_beverly_rcs_rustplotlib.py}
  \label{fig:beverly-head-rcs}
\end{figure}

The head echo RCS measured with the Beverly receiver is shown in
Figure~\ref{fig:beverly-head-rcs}. The RCS is around 20--30 dBsm during the strongest part of the head echo between 90 and 70 km, comparable to or larger than representative microwave-frequency RCS values for large commercial aircraft \citep{skolnik_2001_radar}. Measurement samples outside that interval are also shown, but these are due to receiver noise. The strongest trail echoes in Figure~\ref{fig:trail-echo-roi} reach SNRs near 20 dB, which gives order-of-magnitude peak bistatic RCS estimates up to approximately 40 dBsm per range--Doppler cell.

\begin{figure}
  \centering
  \includegraphics[width=\linewidth]{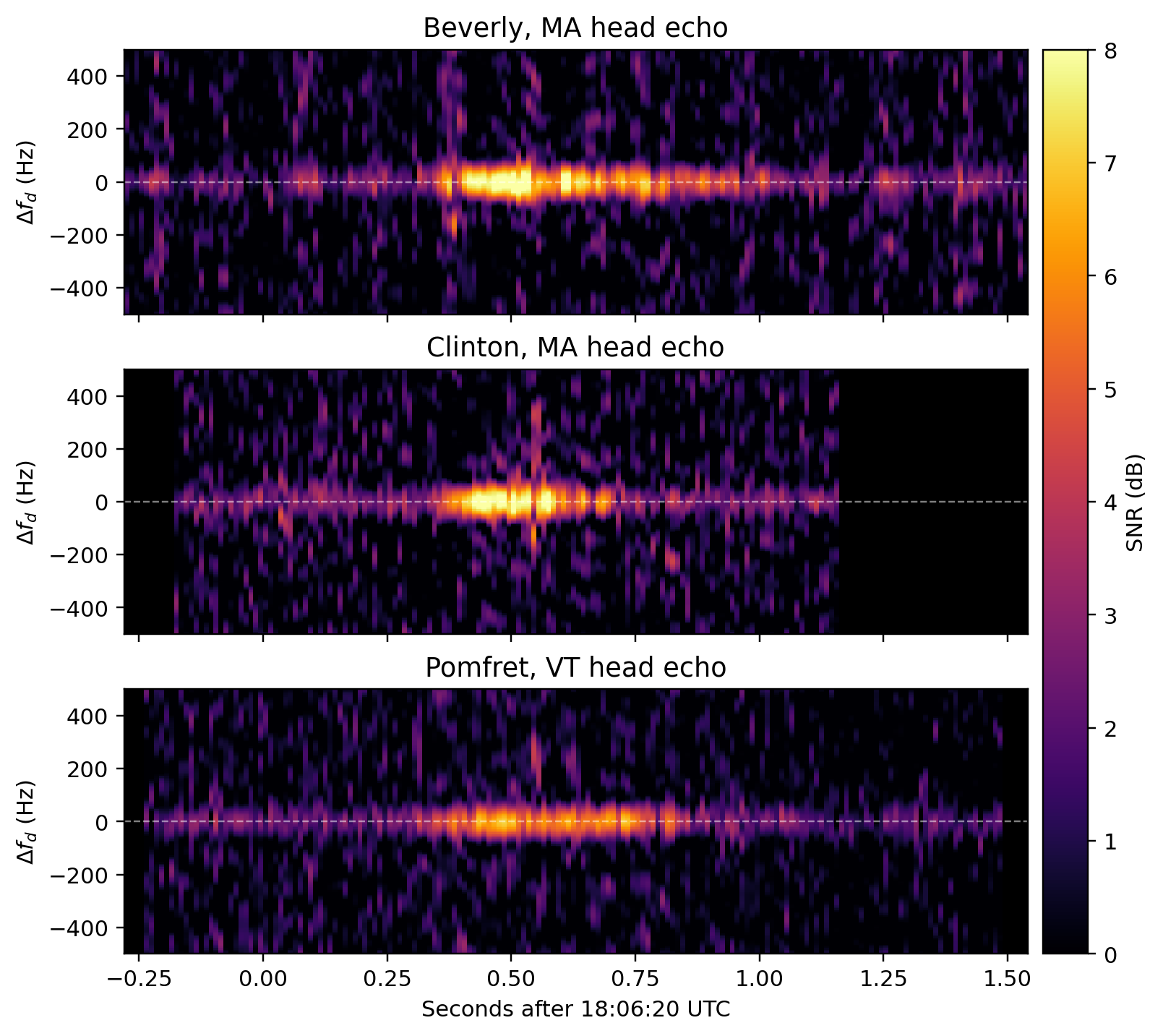}
  \caption{Head-echo Doppler spectra at Beverly, Massachusetts; Clinton,
  Massachusetts; and Pomfret, Vermont.
  The Doppler offset \(\Delta f_d\) is relative to the Gaussian-fitted
  head-echo Doppler; the dashed horizontal lines mark zero offset.
  Matched filtering used one 10 ms code per coherent integration, with
  two spectra averaged incoherently and 16-fold Doppler zero padding. The displayed SNR uses the median spectral power at each
  station as the background estimate.}
  \figsource{reports/plot_head_microdoppler_article.py}
  \label{fig:microdoppler}
\end{figure}

The head-echo Micro-Doppler spectra at Beverly, Clinton, and Pomfret are shown
in Figure~\ref{fig:microdoppler}. The Doppler shift for all three links is shifted to appear at zero, showing only the deviation to a point-like target. During the strongest returns, the spectra contain a single narrow ridge around the fitted head-echo Doppler, with no
resolved secondary tracks indicative of differential fragment velocities.

\begin{figure}
  \centering
  \begingroup
  \normalsize
  \dimen0=\linewidth
  \advance\dimen0 by -164.600000mm
  \ifdim\dimen0<0pt
    \PackageError{rustplotlib}{Article linewidth changed}{Regenerate this figure using the new ArticleLayout.}
  \fi
  \dimen0=\csname f@size\endcsname pt
  \advance\dimen0 by -10.000000pt
  \ifdim\dimen0<0pt \dimen0=-\dimen0\fi
  \ifdim\dimen0>0.01pt
    \PackageError{rustplotlib}{Article body font changed}{Regenerate this figure using the new ArticleLayout.}
  \fi
  \includegraphics[width=164.600000mm]{\detokenize{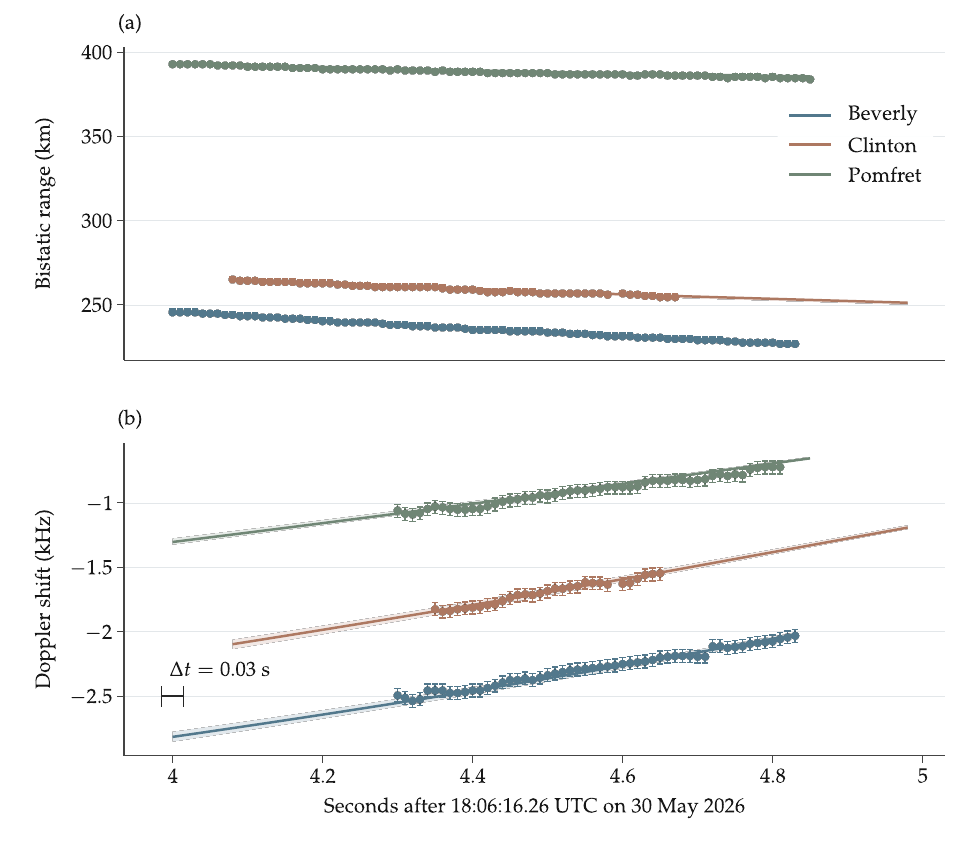}}
  \endgroup
  \caption{Head-echo range--Doppler measurements and best-fit ballistic
  observables.  Points with $\pm2\sigma$ error bars show the
  Beverly, Clinton, and Pomfret head-echo range and Doppler measurements from
  the trajectory-steered, polarization-matched matched
  filter. Three 10 ms code lengths are incoherently averaged together. The measurement correlation length of 0.03 s is indicated in the bottom panel. Solid curves show the corresponding observables predicted by the
  combined head/trail ballistic fit; shaded bands show the 95\% uncertainty
  from the correlated fit covariance.}
  \figsource{reports/replot_article_range_doppler.py; fit_ballistic_head_trail_correlated_appendix.py; reports/export_range_doppler_figure.py; reports/rust_range_doppler/src/main.rs}
  \label{fig:head-range-doppler-fit}
\end{figure}

\begin{figure}
  \centering
  \includegraphics[width=\linewidth]{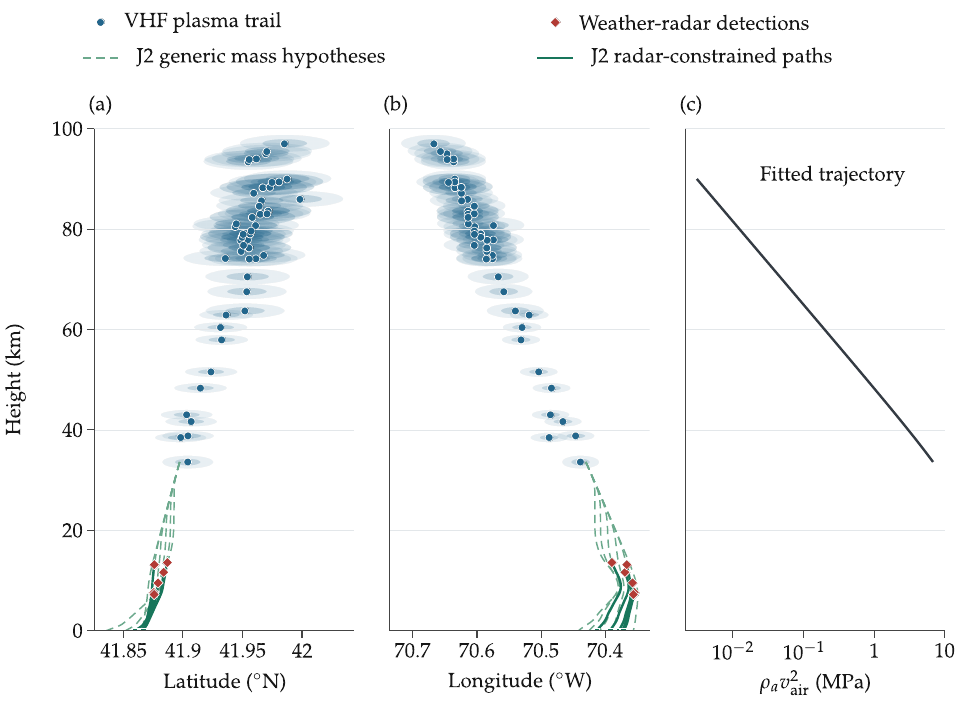}
  \caption{Comparison of the measured
  VHF plasma-trail positions with the dark-flight analysis.  The first two
  panels show latitude and longitude versus altitude.  Markers and translucent
  ellipses show the accepted Beverly sparse-component trail positions and
  fitted component extents. Dashed green curves show the
  Jörmungandr2 generic dark-flight mass hypotheses from
  0.1 g to 10 kg, propagated
  from the model terminus to sea level.  Solid green curves show the six
  radar-constrained dark-flight solutions, with green squares marking their
  modeled endpoints; red diamonds show the KBOX, KENX, and KOKX weather-radar
  detections.  The third panel shows dynamic pressure along the fitted
  trajectory for a 1.6 m diameter spherical body with mass 5000 kg and \(C_D=1\),
  down to 33.61 km using \(p=\rho_a v_{\mathrm{air}}^2\).
  The atmospheric density is from NRLMSIS 2.1.}
  \figsource{reports/plot_fragment_size_hypotheses.py; reports/export_fragment_figure.py; reports/rust_fragments/src/main.rs}
  \label{fig:fragment-size-hypotheses}
\end{figure}

The head-echo measurement set consists of transmit-target-receiver path lengths and rate-of-change of path length measurements for the three stations that measured a head echo. The measurements and the ballistic trajectory fit model obtained by fitting the positions of the trail echoes are shown in Figure~\ref{fig:head-range-doppler-fit}. The measured trail-echo positions and their relationship to the subsequent dark-flight analysis are shown in Figure~\ref{fig:fragment-size-hypotheses}.

\begin{table}
  \centering
  \caption{Initial GCRS state used for orbit propagation at 2026-05-30 18:06:16.260000 UTC (MJD2000 UTC 9646.25435486), 4 s before the first retained head echo.  State order is $(x,y,z,v_x,v_y,v_z)$ with SI units $(\mathrm{m},\mathrm{m},\mathrm{m},\mathrm{m\,s^{-1}},\mathrm{m\,s^{-1}},\mathrm{m\,s^{-1}})$; covariance entries use the corresponding unit products.}
  \label{tab:ballistic-initial-state}
  \begingroup
  \small
  \setlength{\tabcolsep}{2.5pt}
  \begin{tabular}{@{}ll|rrrrrr@{}}
    \toprule
     & Mean / $10^4$ & \multicolumn{6}{c@{}}{Covariance / $10^4$} \\
    \midrule
    Variable & Estimate & $x$ & $y$ & $z$ & $v_x$ & $v_y$ & $v_z$ \\
    \midrule
    $x$ & $10.183$ & $25.84$ & $6.168$ & $3.129$ & $-4.214$ & $0.5103$ & $0.6241$ \\
    $y$ & $487.82$ & $6.168$ & $19.05$ & $16.55$ & $-1.608$ & $-5.345$ & $-5.918$ \\
    $z$ & $437.83$ & $3.129$ & $16.55$ & $39.68$ & $-0.4369$ & $-5.837$ & $-10.19$ \\
    $v_x$ & $-0.87827$ & $-4.214$ & $-1.608$ & $-0.4369$ & $0.8368$ & $0.1308$ & $-0.07903$ \\
    $v_y$ & $-1.588$ & $0.5103$ & $-5.345$ & $-5.837$ & $0.1308$ & $1.788$ & $2.072$ \\
    $v_z$ & $-1.8737$ & $0.6241$ & $-5.918$ & $-10.19$ & $-0.07903$ & $2.072$ & $3.116$ \\
    \bottomrule
  \end{tabular}
  \endgroup
\end{table}

The tri-static head-echo time-of-flight solution and the transmit
interferometry provide complementary estimates of the meteoroid position.  In
the former, the position is constrained by simultaneous bistatic path lengths
from multiple receivers; in the latter, the measured transmit AoD fixes a ray
from Haystack and the bistatic path length fixes the distance along that ray.
For the head-echo samples that passed the AoD selection, the
straight-line fit to the AoD plus delay positions was separated from the
tri-static trajectory by \AodTriFitRmsKm{} km RMS.  The median pointwise
separation of individual AoD plus delay positions from the tri-static
trajectory was \AodTriMedianPointKm{} km.  

\subsection{Weather-Radar Detections and Dark Flight}
\label{sec:weather-radar-dark-flight}

Signatures of falling meteorites from this event are visible in data from
WSR-88D weather radars in the National Oceanic and Atmospheric Administration
(NOAA) NEXRAD radar network.  The signatures are identifiable as such through
a combination of factors: they appear at the time and place of a bolide, the
ensemble of signatures appears as a vertical column consistent with material
falling over a range of altitudes that include values higher than the ambient
weather, and the earliest-arriving signatures include pronounced short-range
turbulence consistent with high-velocity objects.  These features have been
used to identify meteorite falls in NEXRAD data both as they occur and in
archived data, for a total of 61 falls as of July 2026, including the recovered
Lake Huron event (M. Fries, personal communication, 20 July 2026)
\citep{fries_fries_2010,jenniskens_2012_sutters_mill,
jenniskens_2019_creston,heck_2020_hamburg,jenniskens_2020_dishchii,
fries_fries_2025,Scamfer2026}.

Meteorites are seen in data from the KBOX (Boston, Massachusetts), KENX
(Albany, New York), and possibly KOKX (New York City, New York) radars.  The
KBOX data are the most numerous, with signatures evident in four radar sweeps
in the 18:03 UTC data suite.  Altitudes range from 13,595 m down to 7,600 m
above mean sea level (AMSL) for the KBOX data.  Signatures are also evident in
data from the TBOS airport weather radar operated by the Federal Aviation
Administration, at 15,178 m and 19,452 m AMSL.  Clouds in the area at the time
of the fall are found at up to approximately 6,500 m AMSL, well below the
signatures of falling meteorites, so the two sources are distinct.

The SIMONe-constrained terminus and weather-radar centroids produce a J2
terminus location of
\(41.89491^\circ\) N, \(70.41082^\circ\) W.  The azimuth of the bolide ground
track and trajectory depression angle are taken from SIMONe data as
\(115.0^\circ\) and \(72^\circ\), respectively.  The SIMONe path and the
weather-radar signatures are in good spatial agreement. Figure \ref{fig:fragment-size-hypotheses} shows the positions of the VHF radar echoes of meteor plasma, weather radar echoes of falling meteorites, and the path hypotheses of the falling meteorites based on the J2 model.

\begin{figure}
  \centering
  \begin{tikzpicture}[x=\linewidth,y=\linewidth]
    \node[anchor=south west,inner sep=0] at (0,0)
      {\includegraphics[width=\linewidth]{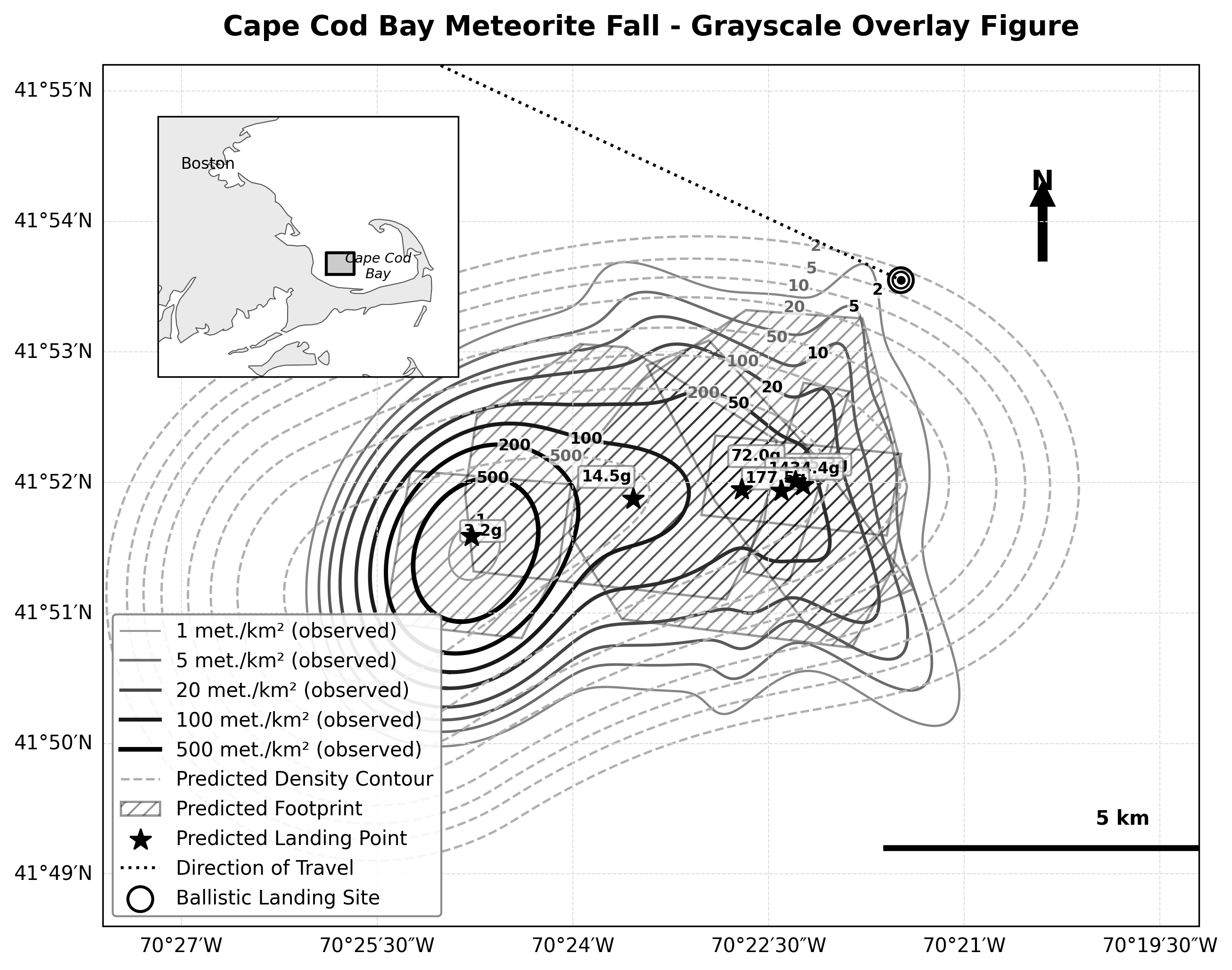}};
    \fill[white] (0.710,0.087) rectangle (0.972,0.140);
    \draw[line width=1.5pt] (0.800,0.100) -- (0.8767,0.100);
    \node[anchor=south,font=\small] at (0.83835,0.105) {1 km};
  \end{tikzpicture}
  \caption{Meteorite landing strewn field
  calculated with the J{\"o}rmungandr2 dark-flight model from weather-radar
  signatures of falling meteorites.  Stars mark the centroids of calculated
  landing sites and are labeled by modeled meteorite mass; hatched polygons
  show the radar-signature footprints translated to their calculated landing
  sites.  Solid isopleths show the calculated number density for the
  radar-observed meteorites and are not mass-weighted, so the more numerous
  western meteorites are also lower mass.  Dashed isopleths show the extended
  calculation for hypothetical meteorites from 0.1 g to 10 kg.  Winds were
  generally from the northeast and displaced falling meteorites toward the
  southwest. Fine protrusions in the contours arise from the polygonal
  radar-signature footprints and should not be interpreted as resolved
  structure in the meteorite distribution.}
  \figsource{Marc Fries: CCB_fig3_2.png; scale-bar overlay: main.tex}
  \label{fig:j2-strewn-field}
\end{figure}

The J2 results indicate that the meteorite strewn field is centered
approximately at \(41.865^\circ\) N, \(70.385^\circ\) W.  The strewn field is
relatively small despite the high terminus altitude because of light winds,
spanning approximately 6 km east--west and 5 km north--south when masses down
to 0.1 g are included.  The six weather-radar-constrained landing solutions
shown in Figure~\ref{fig:j2-strewn-field} have modeled masses from 3.2 g to
1.425 kg.  This does not define the complete range of meteorite masses in the
fall, because larger masses tend to descend rapidly and can escape detection
by the slow-turning, mechanically driven WSR-88D radars.  J2 also calculates
an extended strewn field using a hypothetical mass range from 0.1 g to 10 kg. Figure~\ref{fig:j2-strewn-field} overlays the weather-radar-constrained landing
solutions and the extended hypothetical-mass calculation.

Assuming 10\% survival, conditional on an iron composition, gives an
illustrative total fall mass of 560 kg based on the NASA
Meteoroid Environment Office pre-atmospheric mass estimate of 5600 kg
\citep{nasa_ares_2026_cape_cod_bay}.  Integrating the radar-derived
mass--number distribution instead gives 82 kg when a
dielectric factor \(\lvert K\rvert^2=1.00\) is assumed, with the caveat that
this calculation includes only meteorites observed by radar and not larger
meteorites that escaped detection.  This radar-reflectivity
method is unverified, and the iron-composition survival assumption is not a
composition determination.  These calculations are presented here only to show
that total fall-mass estimates vary widely and require further method
development.  Using the
conservative 82 kg estimate, the modeled ocean-surface density ranges from
approximately 1,000 meteorites km\(^{-2}\) for the approximately 2 g mass range
to 29 meteorites km\(^{-2}\) for 985 g meteorites.  Water depth at the fall site is
approximately 34 m, so horizontal translation during descent to the seafloor
should be negligible compared with the size of the strewn field.

\subsection{Orbit}

\begin{figure}
  \centering
  \begingroup
  \normalsize
  \dimen0=\linewidth
  \advance\dimen0 by -164.600000mm
  \ifdim\dimen0<0pt
    \PackageError{rustplotlib}{Article linewidth changed}{Regenerate this figure using the new ArticleLayout.}
  \fi
  \dimen0=\csname f@size\endcsname pt
  \advance\dimen0 by -10.000000pt
  \ifdim\dimen0<0pt \dimen0=-\dimen0\fi
  \ifdim\dimen0>0.01pt
    \PackageError{rustplotlib}{Article body font changed}{Regenerate this figure using the new ArticleLayout.}
  \fi
  \includegraphics[width=115.220000mm]{\detokenize{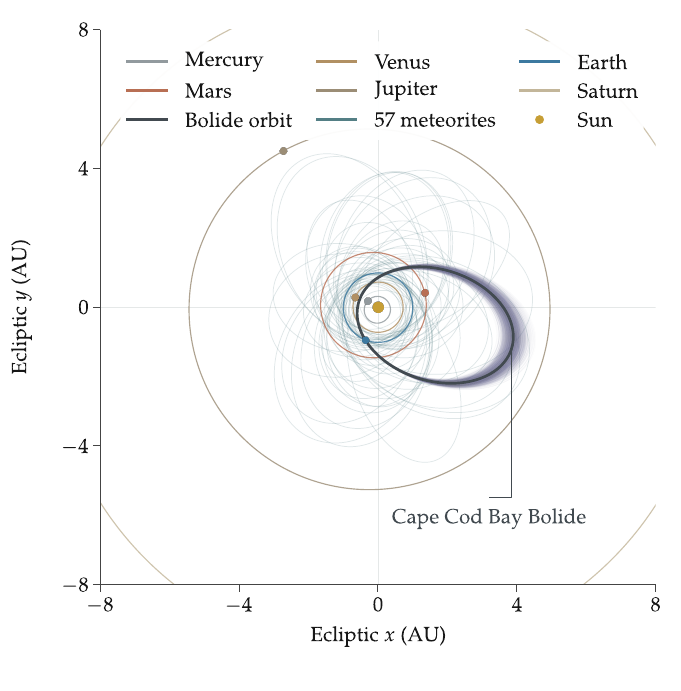}}
  \endgroup
  \caption{Bolide orbit distribution in the J2000 heliocentric ecliptic
  \(x\)--\(y\) plane. The solid dark curve shows the nominal bolide orbit,
  with a leader anchored at aphelion.
  The gray cloud shows \OrbitSampleCount{} Monte Carlo
  orbit samples from the current v\OrbitTrajectoryFitVersion{} manuscript
  orbital-element distribution in Table~\ref{tab:orbit-parameters}.  The 
  pale blue-gray curves show a representative subset of 57 recovered-meteorite orbits
  with complete orbital elements in the Meteorite Orbits compilation
  \citep{meteorite_orbits_info}. Thin colored curves show planetary orbits,
  with circles marking planetary positions at the event epoch.}
  \figsource{reports/export_orbit_figure.py; plot_orbit_xy.py; reports/rust_orbits/src/main.rs}
  \label{fig:orbit-xy}
\end{figure}

Figure~\ref{fig:orbit-xy} shows the estimated orbit based on the VHF radar path. The estimated pre-Earth-encounter orbital elements are given in Table~\ref{tab:orbit-parameters}.  The uncertainty is Monte-Carlo sampled from the initial state vector uncertainty covariance matrix. For comparison, the orbit plot shows a representative subset of 57 recovered-meteorite orbits with complete orbital elements from the meteorite orbits compilation of Meier \citep{meteorite_orbits_info}.

\begin{table}
  \centering
  \caption{Heliocentric mean-ecliptic zenithal attraction corrected orbital elements.}
  \small
\begin{tabular}{llrl}
\toprule
Quantity & Symbol & Unit & Value \\
\midrule
Epoch & & UTC & 2026-05-30 00:53:16.260  \\
Epoch & & MJD & 61190.03699375  \\
Semimajor axis & $a$ & AU & 2.44 $\pm$ 0.10 \\
Perihelion distance & $q$ & AU & 0.6045 $\pm$ 0.0030 \\
Aphelion distance & $Q$ & AU & 4.27 $\pm$ 0.20 \\
Eccentricity & $e$ &  & 0.752 $\pm$ 0.010 \\
Inclination & $i$ & deg & 19.21 $\pm$ 0.18 \\
Argument of perihelion & $\omega$ & deg & 93.4 $\pm$ 0.6 \\
Longitude of ascending node & $\Omega$ & deg & 68.99209 $\pm$ 0.00016 \\
True anomaly at epoch & $\nu$ & deg & 85.9 $\pm$ 0.6 \\
Jupiter Tisserand parameter & $T_J$ &  & 2.99 $\pm$ 0.08 \\
Corrected radiant right ascension & RA & deg & 59.12 $\pm$ 0.34 \\
Corrected radiant declination & Dec & deg & 45.85 $\pm$ 0.18 \\
\hline
Time at 100 km & $t_{100}$ & UTC & 2026-05-30 18:06:19.819  \\
Geodetic latitude at 100 km & $\phi_{100}$ & deg & 41.99 \\
Geodetic longitude at 100 km & $\lambda_{100}$ & deg & -70.69 \\
Entry azimuth at 100 km & Az$_{100}$ & deg & 111.4 \\
Entry elevation at 100 km & El$_{100}$ & deg & -70.2 \\
Ground-relative speed at 100 km & $v_{100}$ & km s$^{-1}$ & 26.00 \\
Terminus altitude & $h_\mathrm{term}$ & km & 33.61 \\
Terminus geodetic latitude & $\phi_\mathrm{term}$ & deg & 41.9105 \\
Terminus geodetic longitude & $\lambda_\mathrm{term}$ & deg & -70.4202 \\
Terminus azimuth, velocity at 100 km & Az$_\mathrm{term}$ & deg & 111.6 \\
Terminus elevation, velocity at 100 km & El$_\mathrm{term}$ & deg & -70.0 \\
\bottomrule
\end{tabular}

  \figsource{reports/generate_orbit_source_table.py}
  \label{tab:orbit-parameters}
\end{table}
\FloatBarrier

\subsection{Infrasound Results}

\begin{figure}[!htbp]
\centering
\includegraphics[width=\linewidth]{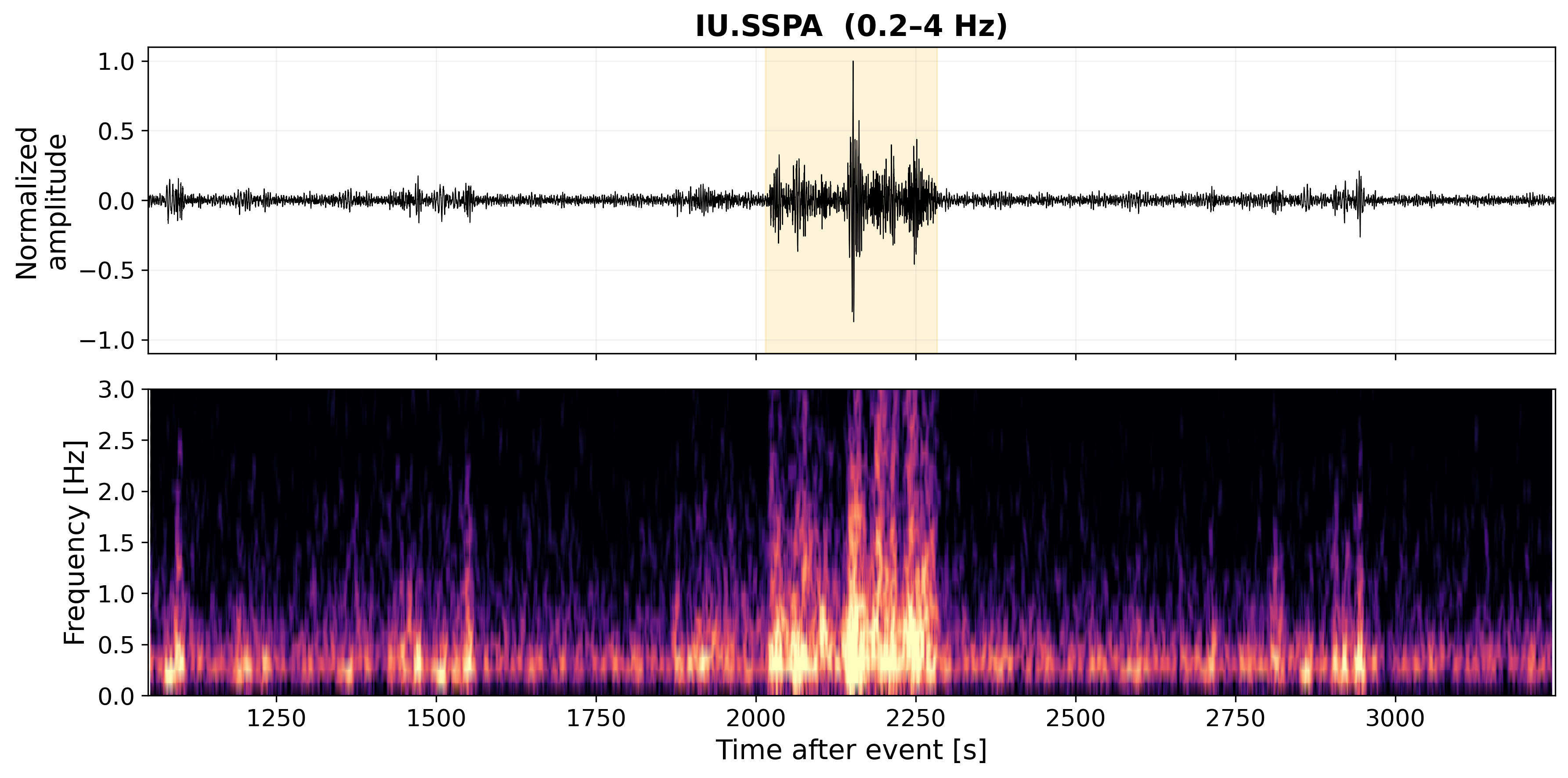}
\caption{Representative single-station infrasound detection at IU.SSPA. Band-pass-filtered waveform and corresponding spectrogram for the 0.2--4 Hz analysis band. The shaded interval denotes the portion of the record identified as the retained bolide signal. The waveform and spectrogram show enhanced low-frequency energy relative to the surrounding background.}
\label{fig:infrasound_signal}
\end{figure}

\begin{figure}[!htbp]
\centering
\includegraphics[width=\textwidth]{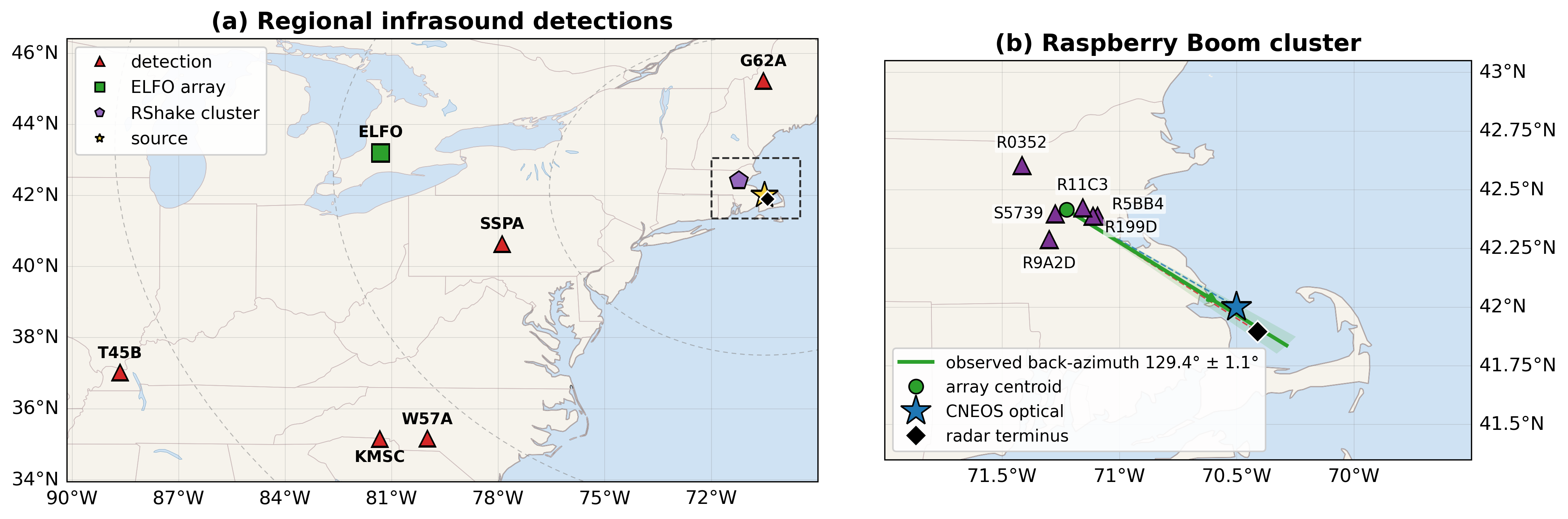}
\caption{(a) Regional distribution of infrasound detections for the 30 May 2026 Cape Cod Bay bolide. The CNEOS peak-brightness location is shown with the VHF radar terminus, the Elginfield Infrasound Array (ELFO), and the Raspberry Shake/Boom cluster. Concentric dashed curves indicate source range. The dashed box marks the area enlarged in panel (b). (b) Directional result from the Raspberry Shake/Boom cluster. The green line and shaded sector show the best-fitting back-azimuth of 129.4$^\circ$ $\pm$ 1.1$^\circ$, derived from inter-station envelope delays under a plane-wave model. The measured direction differs by 2.0$^\circ$ from the bearing to the CNEOS location and by approximately 1.0$^\circ$ from the bearing to the VHF radar terminus.}
\label{fig:infrasound_map}
\end{figure}

Infrasound detections were retained at five single-sensor stations located 358--1649 km from the source. In addition, two geographically and instrumentally independent multi-sensor systems resolved the signal direction. A representative retained arrival at SSPA is shown in Figure~\ref{fig:infrasound_signal}, illustrating the waveform and associated spectral energy. The spatial distribution of detections, ELFO, and the Raspberry Shake/Boom cluster is shown in Figure~\ref{fig:infrasound_map}(a).

\begin{figure}[!htbp]
\centering
\includegraphics[width=0.82\linewidth]{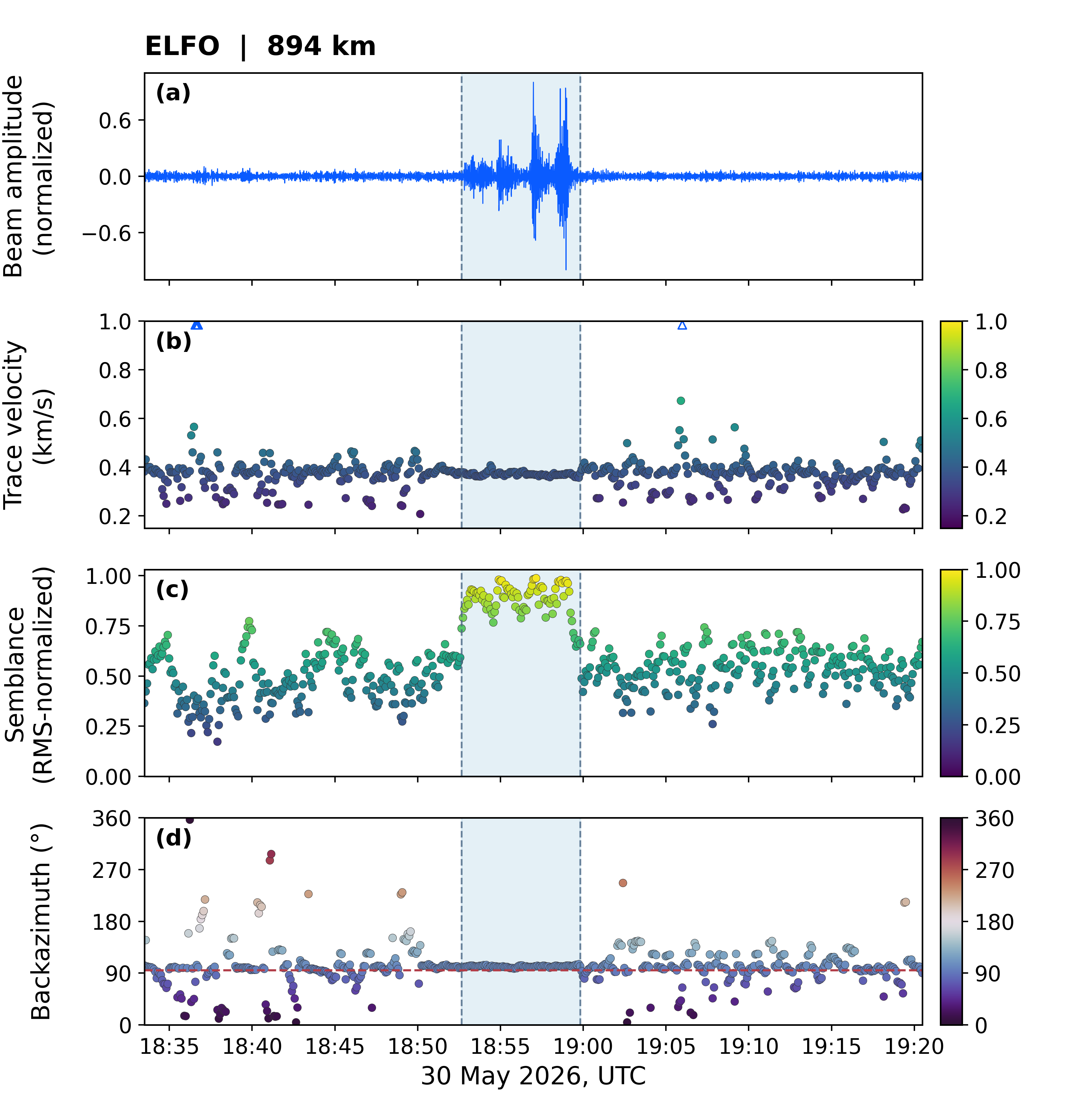}
\caption{ELFO array processing for the Cape Cod Bay bolide at a source range of 894 km. (a) Normalized beam waveform, (b) FK-derived apparent trace velocity, (c) root-mean-square-normalized time-domain semblance, and (d) FK-derived back-azimuth. The shaded interval marks the coherent infrasonic arrival. The horizontal dashed line in panel (d) indicates the bearing toward the source region.}
\label{fig:elfo_beamforming}
\end{figure}

Analysis of the ELFO array data at a range of 894 km identified a coherent infrasonic wavetrain with a stable apparent trace velocity near 0.37 km/s and a back-azimuth of 101.4$^\circ$ $\pm$ 5.3$^\circ$ (Figure~\ref{fig:elfo_beamforming}). The observed back-azimuth differs by +6.6$^\circ$ from the bearing to the CNEOS peak-brightness location and by +6.0$^\circ$ from the bearing to the VHF radar terminus. An independent cross-correlation delay solution gave a back-azimuth of 100.8$^\circ$. The magnitude of the ELFO residual is within the range of directional deviations previously observed for bolide infrasound propagating over regional to long distances, where atmospheric winds and propagation structure can modify the apparent arrival direction \citep{Silber2024}.

The nearby citizen-operated Raspberry Shake/Boom cluster provided an independent short-range directional measurement. The six sensors gave a back-azimuth of 129.4$^\circ$ $\pm$ 1.1$^\circ$, differing by +2.0$^\circ$ from the bearing to the CNEOS peak-brightness location and by approximately -1.0$^\circ$ from the bearing to the VHF radar terminus. The cluster geometry and recovered source direction are shown in Figure~\ref{fig:infrasound_map}(b). Although none of the six sensors constituted a statistically significant detection individually, their joint envelope-coherence measurement was highly significant, with the observed array statistic exceeding all 2000 randomized time-shift realizations and all 728 off-source windows ($p<0.001$).

A first-order assessment of the source-region stratospheric winds placed all retained detections along downwind or crosswind propagation paths, with no upwind detections. The network-averaged dominant period for the five retained single-sensor detections and the ELFO beam was 2.92 $\pm$ 0.41 s. Applying the multi-station bolide period-yield relation of \citet{Silber2025} gives an infrasonic energy estimate of 0.45 kt TNT equivalent. Using the five single-sensor detections alone gives a mean period of 3.02 s and a corresponding energy estimate of 0.51 kt TNT equivalent.

\FloatBarrier

\section{Discussion}

\subsection{Radar Signatures}

Several aspects of the meteor-radar observations support relatively high
mechanical strength of the meteoroid.
Despite its large inferred diameter and mass (1.6 m and 5600 kg, estimated by
NASA's Meteoroid Environment Office \citep{nasa_ares_2026_cape_cod_bay}),
the meteoroid produced a detectable head echo only below approximately 90 km.
By comparison, the much smaller meteoroid observed by
\citet{obenberger2026first} with a similar meteor radar system was detectable
at about 120 km. The entry speeds differed substantially: approximately
25 km s$^{-1}$ for the present event and 72 km s$^{-1}$ for the
\citet{obenberger2026first} event, which makes a direct comparison harder.

For fixed cross-sectional area and drag coefficient, the
kinetic-energy loss rate scales as \(\rho_a v_{\mathrm{air}}^3\). Equal
loss rates therefore require an atmospheric density approximately 24 times
larger at 25 km s$^{-1}$ than at 72 km s$^{-1}$. The MSIS atmospheric
density ratio between 90 and 120 km is approximately 170, substantially larger
than the factor required to compensate for the lower speed in this scaling.
Entry speed and neutral density alone therefore do not explain
the low detection altitude. Head-echo detectability also depends on
composition-dependent ablation and the speed- and species-dependent efficiency
of collisional ionization \citep{vondrak_2008_meteoric_ablation}, as well as
entry angle. A less volatile composition could contribute to the lower initial detection altitude for the head echo.

The micro-Doppler spectra (Figure~\ref{fig:microdoppler}) show no resolved secondary tracks indicative of fragmentation
during the head-echo observation, in contrast
to the clear fragmentation signatures reported by
\citet{obenberger2026first}. This again supports a high mechanical strength meteoroid.

The modeled dynamic pressure reached approximately 7 MPa near the radar
terminus (Figure~\ref{fig:fragment-size-hypotheses}c), using
\(p=\rho_a v_{\mathrm{air}}^2\) following \citet{borovicka_2020_strengths}.
This pressure exceeds the
0.9--5 MPa second-stage fragmentation pressures reported
for ordinary-chondritic meteorite falls by \citet{borovicka_2020_strengths}.

Together with the weather-radar evidence for surviving fragments, these
observations support relatively high mechanical strength of the meteoroid
and the presence of material capable of withstanding substantial aerodynamic
loading.

The effective bistatic head-echo RCS is 100-1000 $\mathrm{m^2}$ (Figure~\ref{fig:beverly-head-rcs}). These values greatly
exceed the projected area of the solid meteoroid and are 100-1000 times larger than the largest RCS reported
for faint head echoes observed with a MST radars with a similar frequency \citep{schult_2013_determination}.
The large return may arise from plasma in the shock layer and near wake of
the moving metre-class meteoroid. The high Doppler shift and agreement with
the modeled range migration show that the dominant scattering plasma most likely is moving
together with the meteoroid. At the upper end, the RCS corresponds to an equivalent
circular diameter of 30--40 m, but this is not a physical plasma diameter:
the estimate is conditional on the transmitted and received polarization,
and the RCS depends on the electron-density distribution and scattering
regime \citep{marshall2017plasma}. At 31.65 MHz, the cold-plasma critical
electron density is approximately \(1.2\times10^{13}~\mathrm{m^{-3}}\).
Simulations of shock-generated plasma surrounding much slower moving hypersonic vehicles and
their near wakes yield electron densities well above this threshold
\citep{esposito2024non}, supporting the possibility of an overdense
shock-layer contribution to the head echo.

The head echo disappears at an altitude of approximately 72 km. The reason
for its disappearance is not entirely clear. This altitude also
coincides with the onset of a promptly generated non-specular trail echo, with
a delay of less than 1 s.  The localized trail positions agree, within their
uncertainties, with the trajectory inferred independently from the tri-static
head-echo ranges and Doppler shifts.  Their short duration and subsonic Doppler
motion favor scattering from transient plasma turbulence in the supersonic
shock wake rather than from a long-lived, kilometre-scale specular trail.  The
relevant flow regime should be examined more carefully; in particular, the
atmospheric mean free path and Knudsen number may ultimately provide a lower
bound on the body size at the onset of shock-generated trail scattering.

Below approximately 70 km altitude, ordinary meteor-train plasma evolution is
not governed by ambipolar diffusion alone.  Ion--molecule chemistry can remove
free electrons through dissociative recombination and attachment, particularly
in dense, low-altitude trails \citep{baggaley_cummack_1974_chemistry,
baggaley_1978_deionization,baggaley_1979_recombination}.  These processes can
shorten radio-echo lifetimes and bias diffusion coefficients inferred from
echo decay times.  The ion chemistry likely explains why the low-altitude trail echo is short lived.  The observed prompt echoes can thus be interpreted as transient, shock-coupled wake
plasma rather than as classical long-lived specular train scatter, although a quantitative
plasma-chemistry model that incorporates the hypervelocity shock front and wake turbulence is still needed to reach further conclusions.

The lack of detectable deceleration is qualitatively consistent with a large
meteoroid, but it does not provide a strong quantitative constraint in this
case.  A sweep over ballistic coefficients indicates that the smallest object
that would avoid detectable deceleration within the observed head-echo segment
would have a mass of only about 0.2 g.  Since the actual object was clearly
much larger than this, the non-detection of deceleration does not provide a
useful lower limit on the meteoroid mass.

\subsection{Weather radar and dark flight modeling}
\label{sec:discussion-weather-radar-dark-flight}

We identify the lowest retained VHF trail echo, at 33.61 km, as the
\emph{radar terminus}: the lower end of the path along which the entry
produces plasma structures detectable by the VHF system. The J2 dark-flight
model assumes that the disappearance of the echo beyond this point results
from breakup into smaller fragments, which decelerate rapidly and are too
small to sustain a sizeable shock-clad plasma structure. In the J2 analysis, the weather-radar meteorite centroids
are back-propagated to a 3 km s\(^{-1}\) transition speed, yielding a terminus
at the same location. Although the radar terminus need not coincide with
an optical endpoint or the time of maximum luminosity, the CNEOS
peak-brightness altitude of 32.0 km \citep{nasa_cneos_2026_fireballs} lies
only about 1.6 km below it.

The radar observations follow the event from the head echo near 90 km,
through prompt shock-wake echoes down to 33.6 km, and then through
hard-target meteorite signatures to 7.6 km above mean sea level. The
weather-radar detections were identified independently of the meteor-radar
observations, whereas the dark-flight model uses the meteor-radar trajectory
and terminus as inputs. Agreement between the detections and the modeled
dark flight therefore supports the reconstructed fall sequence, although it
does not provide a fully independent validation of the meteor-radar trajectory.
Together with the CNEOS optical detection and the infrasound observation,
these measurements provide a consistent account of atmospheric entry and
the subsequent fall of surviving fragments.

The modeled masses range from
3.2 g to 1.425 kg, with tens to hundreds of grams represented prominently;
this is atypical compared with the more usual 1--100 g range.  The total
meteorite observation time is also anomalously short, only 2 min 2 s rather
than the more typical 8--12 min, indicating that small meteorites are scarce
in this fall.  J2 also calculates the slope of log mass versus log \(N\), where
\(N\) is the number of meteorites in a given signature estimated from total
reflected radar energy.  The slope of a linear fit to these data measures the
prevalence of small meteorites relative to large ones and is a proxy for
parent-body mechanical strength.  The Cape Cod Bay fall exhibits the flattest
slope seen to date, \(-1.024\), indicating a mechanically robust body that
survived the bolide with a large population of relatively large meteorites.
The radar and dark-flight evidence are consistent with a dense, mechanically
strong meteoroid, possibly iron-rich, but the composition remains to be
confirmed by recovered material or independent compositional constraints.

\subsection{Source Association}
\label{sec:source-association}

The Jopek--Williams sporadic-meteoroid classification criteria
\cite{jopek_williams_2013} give conflicting dynamical classifications.
Of the 2000 sampled orbits, 63.3\%, 100.0\%, and 97.3\% satisfy the cometary
side of the \(T_J-i\), \(K-i\), and \(P-i\) criteria, respectively,
whereas 94.7\% and 100.0\% satisfy the asteroidal side of the \(Q-i\) and
\(\varepsilon-i\) criteria. The Jupiter Tisserand parameter,
\(T_J=2.99\pm0.08\), lies near the classification boundary of 3.
These fractions describe the sampled orbits relative to empirical boundaries
and do not provide probabilities of cometary or asteroidal origin.
The orbital elements fall within the range reported for recovered meteorites
by \citet{jenniskens_devillepoix_2025,meteorite_orbits_info}. The nominal
semi-major axis exceeds those of 55 of the 75 meteorite orbits (73\%)
listed in Table~2 of
\citet{jenniskens_devillepoix_2025}.

\citet{shober_2021_main_belt} identified numerous sporadic meteoroids on
Jupiter-family-comet-like orbits whose atmospheric penetration indicated
relatively durable material. Their strength classification used the empirical
\(P_E\) criterion, which accounts for luminous-endpoint atmospheric density,
initial mass, entry speed, and entry angle. They proposed that these objects
originated in the main asteroid belt and evolved onto comet-like orbits.
Cape Cod Bay similarly combines an orbit near the cometary--asteroidal
classification boundary with evidence for mechanically resistant material.

A catalogue-neighbour search against Minor Planet Center near-Earth-object and
comet orbit catalogues \cite{Williams2015,MPC2026} and established IAU Meteor
Data Center showers \citep{IAUMDC2026,JENNISKENS2020104821,jenniskens_2023_atlas} shows no convincing shower, comet, or near Earth asteroid association. The closest catalogue neighbours under the Jopek \(D_{\mathrm{H}}\) orbital
dissimilarity criterion \cite{jopek_1993} are near-Earth asteroids rather than
comets.  The closest near-Earth asteroid is \OrbitTopNeaName{} with
\(D_{\mathrm{H}}=\OrbitTopNeaDH{}\), whereas the closest comet is
\OrbitTopCometName{} with \(D_{\mathrm{H}}=\OrbitTopCometDH{}\).  The closest
established shower, the \OrbitTopStreamName{}, has
\(D_{\mathrm{H}}=\OrbitTopStreamDH{}\) and a corrected-radiant separation of
about \(\OrbitTopStreamRadiantSepDeg{}^\circ\), making it a poor association to
any known active stream.  %
%
%
%
%

\if 0 
\subsection{Measurement Considerations}

The event demonstrates the value of retaining raw complex coded-CW voltages
and reprocessing them with a range--Doppler matched-filter bank rather than
relying only on routine meteor-radar detections.  The high-Doppler head echo
can otherwise be masked by subsonic trail or clutter power.  Doppler-gated and
Doppler-whitened products separate these components, while retaining the
complex matched-filter output at the selected Doppler bin preserves the phase
information needed for transmit-AoD estimation.

The head and trail measurements also require distinct physical
interpretations.  Head-echo ranges and Doppler shifts track the moving
meteoroid, whereas the non-specular echoes localize plasma structures left in
its wake and therefore constrain the path rather than the instantaneous body
position.  Displaced or patchy wake scattering naturally produces some
geographic and vertical scatter.  In addition, the 3 km coded-CW range
resolution maps into kilometre-scale position uncertainty when combined with
the AoD measurements; this resolution limit is included when assessing the
agreement between the trail positions and the independently fitted head-echo
trajectory.
\fi

\subsection{Infrasound Source Interpretation}
Bolides generate infrasound through the continuous ballistic shock associated with hypersonic flight and through localized energy release during fragmentation. Gross fragmentation can produce a strong, quasi-spherical shock from a confined portion of the trajectory, distinct from the cylindrical shock generated continuously along the flight path \citep{ReVelle1976,Silber2018,TrigoRodriguez2021,SilberEarthgrazer2026}. As these disturbances decay into infrasound, atmospheric temperature and wind structure control refraction and ducting through tropospheric, stratospheric, and elevated acoustic waveguides over regional to long distances \citep{Drob2010,Albert2023,Silber2024,RonacGiannoneSilber2026}.

The source region is constrained independently by radar, space-based optical observations, and infrasound. CNEOS places peak brightness at 32.0 km altitude, only 1.6 km below the 33.61 km VHF radar terminus, after which weather radar detected surviving fragments in dark flight. The Raspberry Shake/Boom back-azimuth differs by only about 1.0$^\circ$ from the bearing to the VHF radar terminus and by 2.0$^\circ$ from the bearing to the CNEOS peak-brightness location. This agreement associates the regional infrasound with the terminal portion of the luminous trajectory. A similar combination of seismoacoustic, weather-radar, and optical constraints was used to resolve trajectory and fragmentation for the 2025 Alaska fireball \citep{Scamfer2026}.

CNEOS coordinates, however, should not be treated as exact ground truth. Event-specific measurement uncertainties are not reported, and comparisons with independently constrained fireballs show event-to-event variability in the reported parameters \citep{Devillepoix2019,PenaAsensio2025,Wilson2025}. The 2.0$^\circ$ acoustic offset is therefore consistent with the CNEOS source location at the precision relevant here. Radar traces the meteoroid and its plasma wake to the end of the detectable entry path and subsequently detects surviving material in dark flight; the space-based observation identifies the strongest optical energy release at nearly the same altitude; and infrasound records the atmospheric shock radiated from that region. These independent measurements favor gross fragmentation near the end of luminous flight as the dominant source of the regional infrasound.

ELFO provides an independent directional constraint from a source-receiver distance of nearly 900 km. Back-azimuth deviations can arise from both source geometry and atmospheric refraction \citep{Silber2025Angle}. The Cape Cod Bay bolide trajectory is steep, for which geometry-induced directional offsets are expected to be substantially smaller than for shallow entries \citep{Silber2025Angle}. The $\sim6^\circ$ ELFO residual is therefore compatible with a contribution from atmospheric winds and propagation structure along the long path \citep{deGrootHedlin2010,Drob2010,Hedlin2013,Averbuch2022,Silber2024}; quantitative separation of source-geometry and propagation effects would require trajectory-dependent ray tracing.

The network-averaged dominant period corresponds to an infrasonic energy estimate of $\sim0.5$ kt using the multi-station bolide period-yield relation of \citet{Silber2025}. The CNEOS estimate is 1.1 kt, and the difference lies within the established dispersion of empirical bolide period-yield scaling \citep{Ens2012,Silber2025}. The two estimates are based on independent observables: the CNEOS energy is inferred from optical radiated energy, whereas the infrasonic estimate is derived from the dominant acoustic period. Interpreted as pre-atmospheric kinetic energy at the radar-derived entry speed of 26.0 km/s, the 0.5 kt infrasonic estimate corresponds to an equivalent spherical diameter of $\sim1.5$ m for a chondritic body or $\sim1.1$ m for iron, using representative meteorite bulk densities \citep{ConsolmagnoBritt1998}.

The Raspberry Shake/Boom result extends previous use of spatially distributed single-sensor stations for bolide localization through multilateration \citep{Wilson2025}. In this case, inter-station envelope timing provides a statistically significant direct back-azimuth even though the individual citizen-operated sensors are inconclusive. Distributed pressure-sensor networks can therefore provide directional constraints without a purpose-built infrasound array when sufficient spatial coverage and timing coherence are available.

The same seismoacoustic principles apply to artificial atmospheric entries and space debris. For example, the Origins, Spectral Interpretation, Resource Identification, and Security-Regolith Explorer (OSIRIS-REx) sample return capsule provided a controlled hypersonic source with a known trajectory for testing infrasound source localization, signal evolution, and atmospheric propagation \citep{SilberEtAl2024OSIRIS,SilberBowman2025}. Recent reconstructions of a Soyuz-2.1b, and Falcon 9 reentries further demonstrates that meteor-derived optical and seismoacoustic methods can be applied to space-debris trajectories \citep{Hatty2026,hupe26}. These applications link natural bolide monitoring with characterization of artificial reentries and other unplanned hypersonic atmospheric events.

\section{Conclusions}

The combined observations demonstrate that meteor-radar plasma tracking can bridge the upper-atmospheric trajectory and the onset of meteorite dark flight even when precise optical meteor camera observations are not available. Together with weather-radar and infrasound measurements, this permits reconstruction of the pre-impact orbit, terminal fragmentation region, plasma production, and subsequent fall.

The weather-radar-inferred fragment mass distribution, the modeled
aerodynamic pressure of approximately 7 MPa near the radar terminus, and
the absence of resolved fragmentation signatures in the head-echo
micro-Doppler spectra together support a mechanically strong meteoroid.

The reconstructed orbit lies near the cometary--asteroidal classification
boundary, with no convincing association with a catalogued parent body or
established meteor shower. This combination of orbital characteristics and
material durability is consistent with the population of sporadic meteoroids
on Jupiter-family-comet-like orbits described by
\citet{shober_2021_main_belt}, who argued that most originated as main-belt
asteroidal fragments whose orbits subsequently evolved into this regime.

The measured trajectory and height-dependent echo strengths and lifetimes,
together with the infrasound energy estimate of approximately 0.5 kt TNT
equivalent and evidence for a source near the radar terminus, provide
constraints for modeling the hypersonic shock flow, its plasma wake, and
acoustic radiation from the flow and terminal fragmentation. Such models
should explain why the head echo disappears near 72 km while prompt,
short-lived non-specular trail echoes remain detectable down to 33.6 km,
and test whether shock-generated ionization, plasma chemistry, and wake
turbulence can account for their rapid growth and decay.

\section*{Data availability}
The source radar data set is archived on Zenodo as the May 30th 2026 New
England Bolide Radar Dataset \cite{volz_2026_20507006}.
Infrasound waveform data used in this study were accessed through the International Federation of Digital Seismograph Networks (FDSN) and are freely available through its data services. We also provide a Zenodo repository with raw infrasound data (DOI: 10.5281/zenodo.22732819). \ifdefined\ArxivVersion The repository will be made public upon acceptance.\else \fi

\begin{acknowledgments}
EAS and SA were supported by the Laboratory Directed Research and Development (LDRD) program (project number 236896) at Sandia National Laboratories, a multimission laboratory managed and operated by National Technology and Engineering Solutions of Sandia, LLC., a wholly owned subsidiary of Honeywell International, Inc., for the U.S. Department of Energy’s National Nuclear Security Administration under contract DE-NA0003525. 

JLC and MC acknowledge support from the European Office of AFOSR 717
(EOARD) grant FA8655-23-1-7017.

RV, PJE, FDL, JS, and AW acknowledge support for MIT Haystack Observatory participation under the NSF grants AGS-2504079 and AGS-1933007.

The authors acknowledge support from Colin Lonsdale, Will Rogers, Aleks Pop Stefanija, Chris McKenney, Steve Holmberg (MIT Haystack Observatory) and George Hassel (Siena University) for radar implementation, technical assistance, and hosting of nodes for the SIMONe Haystack multistatic meteor radar system.

Sandia National Laboratories is a multi-mission laboratory managed and operated by National Technology \& Engineering Solutions of Sandia, LLC (NTESS), a wholly owned subsidiary of Honeywell International Inc., for the U.S. Department of Energy’s National Nuclear Security Administration (DOE/NNSA) under contract DE-NA0003525. This article has been authored by an employee of National Technology \& Engineering Solutions of Sandia, LLC under Contract No. DE-NA0003525 with the U.S. Department of Energy (DOE). The employee owns all right, title and interest in and to the article and is solely responsible for its contents. The United States Government retains and the publisher, by accepting the article for publication, acknowledges that the United States Government retains a non-exclusive, paid-up, irrevocable, world-wide license to publish or reproduce the published form of this article or allow others to do so, for United States Government purposes. The DOE will provide public access to these results of federally sponsored research in accordance with the DOE Public Access Plan \url{https://www.energy.gov/downloads/doe-public-access-plan}. This paper describes objective technical results and analysis. Any subjective views or opinions that might be expressed in the paper do not necessarily represent the views of the U.S. Department of Energy or the United States Government. 

OpenAI Codex (\url{https://openai.com/codex/}) assisted with manuscript
editing and the development of data analysis and figure-generation software.
\end{acknowledgments}

\clearpage

\bibliographystyle{aasjournalv7.1}
\bibliography{reports/pss_references,reports/infrasound}

%
\iffalse
\appendix
\section{Supporting Memo Index}
\label{app:supporting-memo-index}

The detailed supporting diagnostics have been moved out of the main manuscript and into standalone memos in this Overleaf repository.  The numbered memo list is:

\begin{enumerate}
  \item \textbf{Radar-cross-section calibration memo}: \path{rcs_memo.tex}.
  \item \textbf{Sparse compressive-sensing subtraction memo}: \path{sparse_compressive_sensing_memo.tex}.
  \item \textbf{Synthetic RDMF measurement-chain memo}: \path{synthetic_rdmf_measurement_memo.tex}.
  \item \textbf{Radar signal-processing memo}: \path{radar_signal_processing_memo.tex}.
  \item \textbf{Trajectory-fitting memo}: \path{trajectory_fitting_memo.tex}.
  \item \textbf{Catalogue-association memo}: \path{catalogue_association_memo.tex}.
  \item \textbf{Appendix diagnostics memo}: \path{appendix_diagnostics_memo.tex}.
  \item \textbf{Bolide terminus-estimation memo}: \path{terminus_estimation_memo.tex}.
  \item \textbf{Fragment-size hypothesis memo}: \path{fragment_size_hypothesis_memo.tex}.
\end{enumerate}
\fi

\end{document}